\documentclass[epj]{svjour}
\usepackage{cite}
\usepackage[pdftex]{color}
\usepackage[latin1]{inputenc}       % allow Latin1 characters
\usepackage{graphicx}
\usepackage{amsmath}
\usepackage{amsfonts}
\usepackage{bbm}
\usepackage{ae}
\usepackage{subfigure}

\newcommand{\bra}[1]{\langle #1|}
\newcommand{\ket}[1]{|#1\rangle}

\newcommand{\mean}[1]{\langle #1 \rangle}

\begin{document}

\title{Anderson Localization in 1D Systems with Correlated Disorder}
\author{Alexander Croy\inst{1}\and Philipp Cain\inst{2}\and Michael Schreiber\inst{2}}
\institute{Max-Planck-Institute for the Physics of Complex Systems,
           N\"{o}thnitzer Str.\ 38, 01187 Dresden, Germany \and
           Technische Universit\"{a}t Chemnitz, Institut f\"{u}r Physik,  
           09107 Chemnitz, Germany}
           
         \abstract{Anderson localization has been a subject of intense
           studies for many years. In this context, we study
           numerically the influence of long-range correlated disorder
           on the localization behavior in one dimensional systems. We
           investigate the localization length and the density of
           states and compare our numerical results with analytical
           predictions. Specifically, we find two distinct
           characteristic behaviors in the vicinity of the band center
           and at the unperturbed band edge, respectively. Furthermore
           we address the effect of the intrinsic short-range
           correlations.}

\maketitle

%%%%%%%%%%%%%%%%%%%%%%%%%%%%%%%%%%%%%%%%%%%%%%%%%%%%%%%%%%%%%%%%%%%%%%%%%%%%%%%%%%%%%%%%%%%%%%%%%%%%%%%%%%%%%%%%%%%%%%%%%%%%%%%%%
%
% CHAPTER: Introduction
%
%%%%%%%%%%%%%%%%%%%%%%%%%%%%%%%%%%%%%%%%%%%%%%%%%%%%%%%%%%%%%%%%%%%%%%%%%%%%%%%%%%%%%%%%%%%%%%%%%%%%%%%%%%%%%%%%%%%%%%%%%%%%%%%%%
%
\section{Introduction}
Over the past fifty years the Anderson model of localization has
become a paradigm for investigations of electronic transport in the
presence of static disorder \cite{And58,KraM93}. Depending on the
strength of the disorder the wave functions of non-interacting
electrons vary from delocalized to exponentially localized.  This so
called Anderson localization was demonstrated recently in experiments
with matter waves of Bose-Einstein condensates in random or
quasi-periodic optical potentials \cite{BilJZB08,RoaDFF08}. Due to
significant electron-electron and electron-phonon interactions in
realistic materials, e.g.  imperfect crystals, the direct observation
of Anderson localization is rather difficult.  Therefore the success
of the model is driven to a large extent by the universal properties
of the resulting phenomena.

The characteristics of the disorder potential have a prominent
influence on Anderson localization \cite{KraM93}.  For one dimensional
(1D) systems with a spatially uncorrelated potential it has been shown
that all electronic states are exponentially localized \cite{GolMP77}.
Later it was realized that by introducing correlations, one can partly
suppress the localization, at least for weak disorder \cite{IzrK99}.
Moreover, there have been discussions of a delocalization-localization
transition even in 1D for long-range correlated disorder potentials
\cite{deML98, ShiNN04, Kay07}. However, it has early been recognized
that the apparent transition results from a rescaling of the disorder
potential by normalizing the variance, which becomes system-size
dependent for the considered self-affine potential landscapes
\cite{KanRBH00}.

In the present work, we focus on long-range power law correlated
potentials with a correlation function $C(\ell) \propto
\ell^{-\alpha}$ and correlation exponents $\alpha > 0$. The variance
of the resulting potential is independent of the system size. This so
called scale-free disorder is by no means artificial. It appears
naturally in a large variety of physical systems
\cite{Isi92,PenBGH92,VidMNM96}.

The outline of this article is as follows. In the next section we
briefly review the main results for Anderson localization in 1D and in
the presence of correlated disorder. We also discuss the numerical
methods used in our calculations. In Sec.\ \ref{sec:results} we
present and discuss the obtained results.  Finally, the article is
summarized in Sec.\ \ref{sec:summary}.

%%%%%%%%%%%%%%%%%%%%%%%%%%%%%%%%%%%%%%%%%%%%%%%%%%%%%%%%%%%%%%%%%%%%%%%%%%%%%%%%%%%%%%%%%%%%%%%%%%%%%%%%%%%%%%%%%%%%%%%%%%%%%%%%%
%
% CHAPTER: Model and Numerical Techniques
%
%%%%%%%%%%%%%%%%%%%%%%%%%%%%%%%%%%%%%%%%%%%%%%%%%%%%%%%%%%%%%%%%%%%%%%%%%%%%%%%%%%%%%%%%%%%%%%%%%%%%%%%%%%%%%%%%%%%%%%%%%%%%%%%%%
%
\section{Model and Numerical Techniques}
\subsection{Anderson Model of Localization with Long-Range Correlated Disorder}
The Anderson model \cite{And58,KraM93} is widely used to investigate
the phenomenon of localization in disordered materials. It is based
upon a tight-binding Hamiltonian in site representation,
\begin{equation}
	\mathcal{H} = \sum_{i} \varepsilon_i \ket{i}\bra{i} 
    					- t\;\sum_{\mean{i\,j}}\,\ket{i}\bra{j}\;,
  	\label{eq:AndersonHam}
\end{equation}
where $\ket{i}$ is a localized state at lattice site $i$ and
$\mean{i\,j}$ denotes a restriction of the sum to $i$ and $j$ being
nearest neighbors. The hopping parameter $t=\hbar^2/(2 m^* a^2)$ is
defined in terms of the lattice spacing $a$ and the effective mass
$m^*$.  In the following, we set $t\equiv 1$ and thus fix the unit of
energy; lengths are measured in units of $a$. The on-site potentials
$\varepsilon_i$ are random numbers, chosen according to some
probability distribution $P(\varepsilon)$ characterized by the mean
$\mean{\varepsilon_i}$ and the correlation function
$\mean{\varepsilon_i \varepsilon_{i+\ell}} = C(\ell)$. However,
usually the site energies are taken to be statistically independent.
For example, a common choice of $P(\varepsilon)$ is a Gaussian white
noise distribution or a box distribution of width $W$, both with
$\mean{\varepsilon_i}=0$ and $C(\ell)=\frac{W^2}{12} \delta_{\ell,0}$.
Other distributions have also been considered
\cite{BulSK87,KraM93,OhtSK99,RomS03}.
In the present work we are interested in the influence of long-range
correlated disorder potentials on the Anderson model, i.e., using a
correlation function of the form
\begin{equation}\label{eq:PLCorrFunc}
  C(\ell) \equiv \mean{\varepsilon_i \varepsilon_{i+\ell}} \propto |\ell|^{-\alpha}
\end{equation}
where $\alpha$ is the correlation exponent which determines the strength of the 
correlations. The associated spectral density is given by
\begin{equation}\label{eq:SpecDens}
	S(q) = \sum_\ell C(\ell) e^{-\imath q \ell} \propto |q|^{\alpha - 1 }\;.
\end{equation}
In order to generate physically reasonable long-ranged correlated
potentials, the correlation function \eqref{eq:PLCorrFunc} should
decay with distance $\ell$ and therefore $\alpha>0$.  For 1D systems
the correlations are considered to be long ranged for $0<\alpha<1$.
The value $\alpha = 1$ corresponds to the case of uncorrelated
disorder. Notice that the delocalization-localization transition
observed in Refs.\ \cite{deML98, ShiNN04, Kay07} relies on power-law
correlations with $\alpha<0$. These correlations increase with
distance $\ell$ and lead to a system-size dependence of the disorder
potential. Thus a rescaling of the results is required which yields
the corrected localization behavior without transition
\cite{KanRBH00}.

In general, it is extremely complicated to obtain analytical results
of transport properties for the Anderson model of localization. For
example, only in the 1D case rigorous proofs of strong
localization for all energies and disorder strengths have been given
\cite{GolMP77}. Moreover, the explicit energy and disorder dependence
of the localization length $\Lambda$ for weak disorder has been
derived \cite{Tho79,PasF92}.  There are also some results for 1D
systems with long-range correlated disorder. For energies close to the
band center ($|E| \ll 2$) a weak disorder expansion for $\Lambda$
yields \cite{IzrK99},
\begin{equation}
	\Lambda\left(q, W\right) = \left[ \frac{1}{8 \sin^2(q)} \frac{W^2}{12} S(2 q) \right]^{-1}\;,
	\label{eq:1DWeakDisExpansion}
\end{equation}
where $E(q) = -2 \cos(q)$. It is important to notice that even for the
correlated case the localization length remains proportional to
$W^{-2}$. Thus, for small disorder strength and in the vicinity of the
band center, the presence of correlations leads to an {\it effective}
disorder strength $W_{\rm eff} = W \sqrt{S(2 q)}$, which depends on
energy via the spectral density $S$.

On the other hand, close to the unperturbed band edge ($|E|=2$) a
different universal behavior can be found \cite{CohES85, Rus02}. Here,
the density of states (DOS) and the localization length are characterized by
units of energy $\epsilon$ and length $\lambda$,
\begin{subequations}\label{eq:BEScaling}
\begin{eqnarray}
	\rho_\alpha(E) 		&=& \lambda^{-1} \epsilon^{-1} \, f_ \alpha( E/\epsilon )\;, \\
	\Lambda_\alpha(E) 	&=& \lambda \, g_ \alpha( E/\epsilon )\;,
\end{eqnarray}
\end{subequations}
where $f_\alpha$ and $g_\alpha$ are universal functions. For
uncorrelated potentials, i.e.\, $\alpha = 1$, there is a closed
analytic form available for both functions \cite{DerG84}. The units
$\epsilon$ and $\lambda$ can be expressed in terms of the hopping
parameter $t$ and the disorder strength $w = W/\sqrt{12}$,
\begin{subequations}
\begin{eqnarray}
	\epsilon &=& w^{4/(4-\alpha)}\, \\
	\lambda  &=& w^{-2/(4-\alpha)}\;.
\end{eqnarray}
\end{subequations}
In particular, it follows that at the unperturbed band edge ($|E|=2$), 
\begin{subequations}	\label{eq:1DPowerDecay}
\begin{eqnarray}
	\Lambda_\alpha( |E|=2, W) &\propto W^{-y}\;,\\
	\rho_\alpha( |E|=2, W) &\propto W^{-y}
\end{eqnarray}
\end{subequations}
with $y = 2/3$ for $\alpha=1$ and $y=2/(4-\alpha)$ for $\alpha \le 1$
\cite{DerG84, RusHW98}. Most interestingly, within the so called
white-noise model (WNM) scaling expressions identical to Eqs.\
\eqref{eq:BEScaling} are obtained where $\alpha$ is replaced by the
dimension $d$ of the system. We would like to emphasize that the WNM
does not consider long-range correlated disorder, but the similarity
suggests that the influence of power-law correlations at the band edge
may possibly be interpreted in terms of an {\it effectively reduced
  dimensionality}.

\subsection{Numerical Methods}\label{sec:Methods}
In order to generate random numbers with power-law correlations 
\eqref{eq:PLCorrFunc} we use the
modified Fourier filtering method ({FFM}) \cite{MakHSS96} with one
additional step. Hereby, the correlation function is given by
\begin{equation}\label{eq:MakseCorrFun}
	C(\ell) = \left(1+\ell^2\right)^{-\alpha/2} \propto |\ell|^{-\alpha} \quad (\ell\gg 1)\;,
\end{equation}
which avoids the singularity at $\ell=0$ and therefore improves the
filtering. For large distances $\ell\gg 1$ this form resembles the
desired power-law behavior \eqref{eq:PLCorrFunc}. An additional
benefit of Eq.\ \eqref{eq:MakseCorrFun} consists in avoiding aliasing
effects, which may obscure the long-range character of the
correlations.  The spectral density $S(q)$ can be calculated
analytically and is given in terms of a modified Bessel function
\cite{MakHSS96}.
Overall, the special choice of the correlation function
\eqref{eq:MakseCorrFun} leads to a modified behavior with respect to
the correlation exponent $\alpha$.  While for the pure power-law the
case of uncorrelated disorder is obtained for $\alpha = 1$ (see also
Eq.\ \eqref{eq:SpecDens}), the modified FFM generates uncorrelated
random numbers only in the limit $\alpha\to\infty$. Therefore, it is
expected to find an influence on the localization behavior even for
$\alpha>1$.

Additionally, we shift and normalize the obtained sequence of
correlated random numbers such that the mean vanishes and the variance
is $W^2/12$ \cite{Kay07}.  Thereby, only the strength of the disorder
is adjusted while the correlations of the potential are preserved.
Then we calculate the localization length using a standard
transfer-matrix method (TMM) \cite{KraM93} with a new seed of the
random number generator for each parameter combination ($E, W,
\alpha$). The DOS is calculated by diagonalizing the Hamiltonian
\eqref{eq:AndersonHam} and counting the eigenvalues in finite energy
bins. In each case $100$ realizations are taken into account for
averaging.

Due to the symmetry of the Hamiltionian (\ref{eq:AndersonHam}) and the
chosen symmetric disorder distribution $P(\varepsilon)$ the results
should depend on the absolute value of $E$ only. This is confirmed by
our calculations.  Therefore we use the arithmetic average of values
for $-E$ and $+E$ to improve the accuracy of our results.

In the following, we consider chains of fixed length $L=2^{19}$ for
the TMM calculations and $L=2^{13}$ for the DOS computation. The
correlation exponents are $\alpha = \infty, 2.0, 1.0$ and $0.5$.

%%%%%%%%%%%%%%%%%%%%%%%%%%%%%%%%%%%%%%%%%%%%%%%%%%%%%%%%%%%%%%%%%%%%%%%%%%%%%%%%%%%%%%%%%%%%%%%%%%%%%%%%%%%%%%%%%%%%%%%%%%%%%%%%%
%
% CHAPTER: Numerical Results and Discussion
%
%%%%%%%%%%%%%%%%%%%%%%%%%%%%%%%%%%%%%%%%%%%%%%%%%%%%%%%%%%%%%%%%%%%%%%%%%%%%%%%%%%%%%%%%%%%%%%%%%%%%%%%%%%%%%%%%%%%%%%%%%%%%%%%%%
%
\section{Results and Discussion}\label{sec:results}
\begin{figure}[bt]
	\center
        \includegraphics[width=\columnwidth]{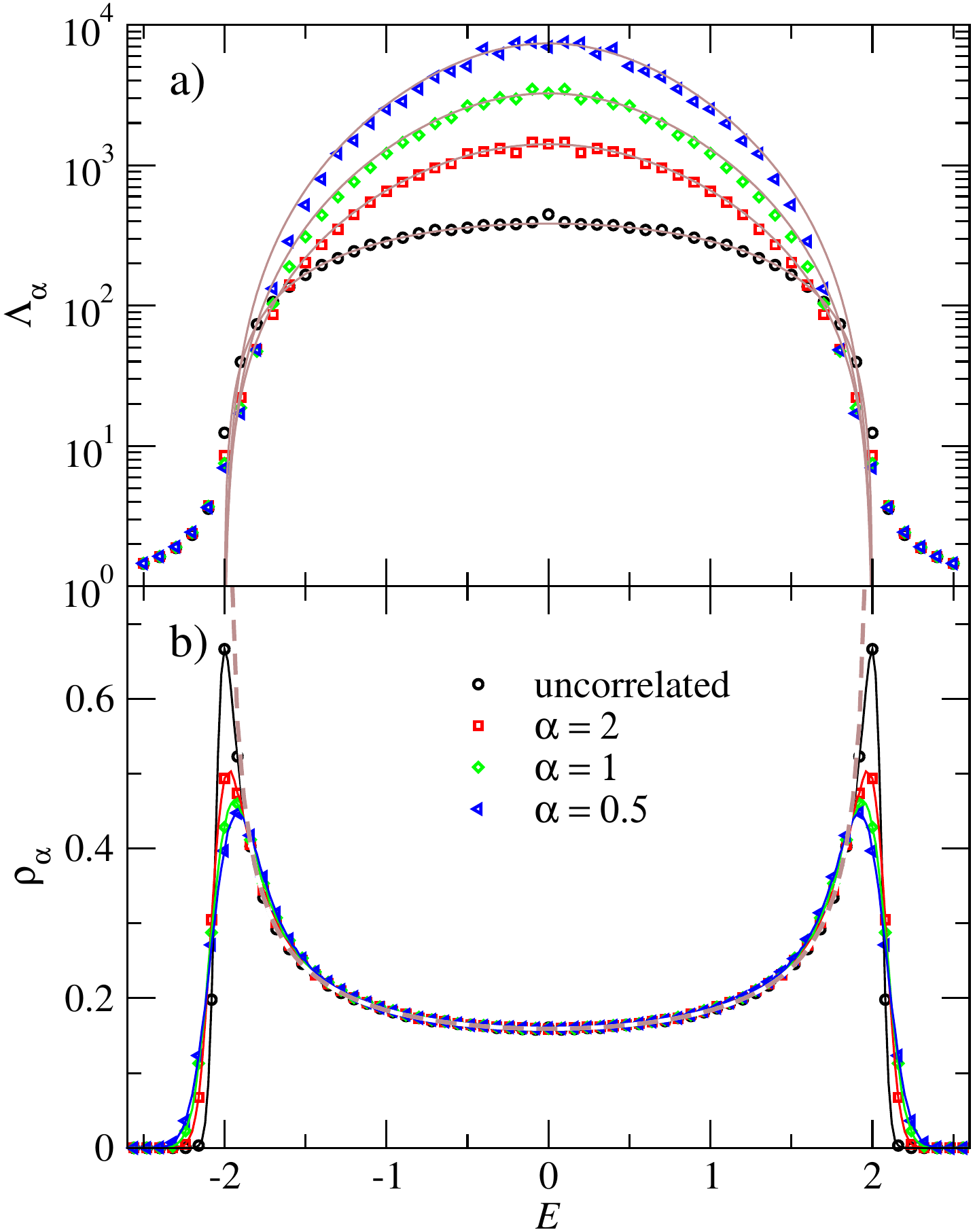}
	\caption{(Color online) Localization length
          $\Lambda_\alpha$ vs energy $E$ at $W=0.5$ for a 1D chain of
          length $L=2^{19}$ obtained from TMM calculations (a).  Full
          lines correspond to $\Lambda_\alpha$ given by Eq.\
          \eqref{eq:1DWeakDisExpansion}. DOS $\rho_\alpha$ for a 1D
          chain of length $L=2^{13}$ (b). The dashed line denotes the exact
          DOS for a 1D system without any disorder, $\rho = 1/(\pi
          \sqrt{4-E^2})$.  Different symbols/colors denote various
          correlation parameters $\alpha$.}
	\label{fig:1DEngLocW05}
\end{figure}

In Fig.\ \ref{fig:1DEngLocW05} we show the energy dependence of
$\Lambda$ and the DOS for fixed disorder strength $W=0.5$ and
different correlation strengths $\alpha$. For energies inside the
unperturbed band ($|E| < 2$) the localization length is increased by
long-range correlations.  In this region one observes an overall good
agreement with the weak-disorder result given by Eq.\
\eqref{eq:1DWeakDisExpansion}.  On the other hand, near the band edge
($|E| \approx 2$) $\Lambda$ is decreased by correlations, which will
be discussed in more detail below.
\begin{figure}[bt]
	\center
	\includegraphics[width=\columnwidth]{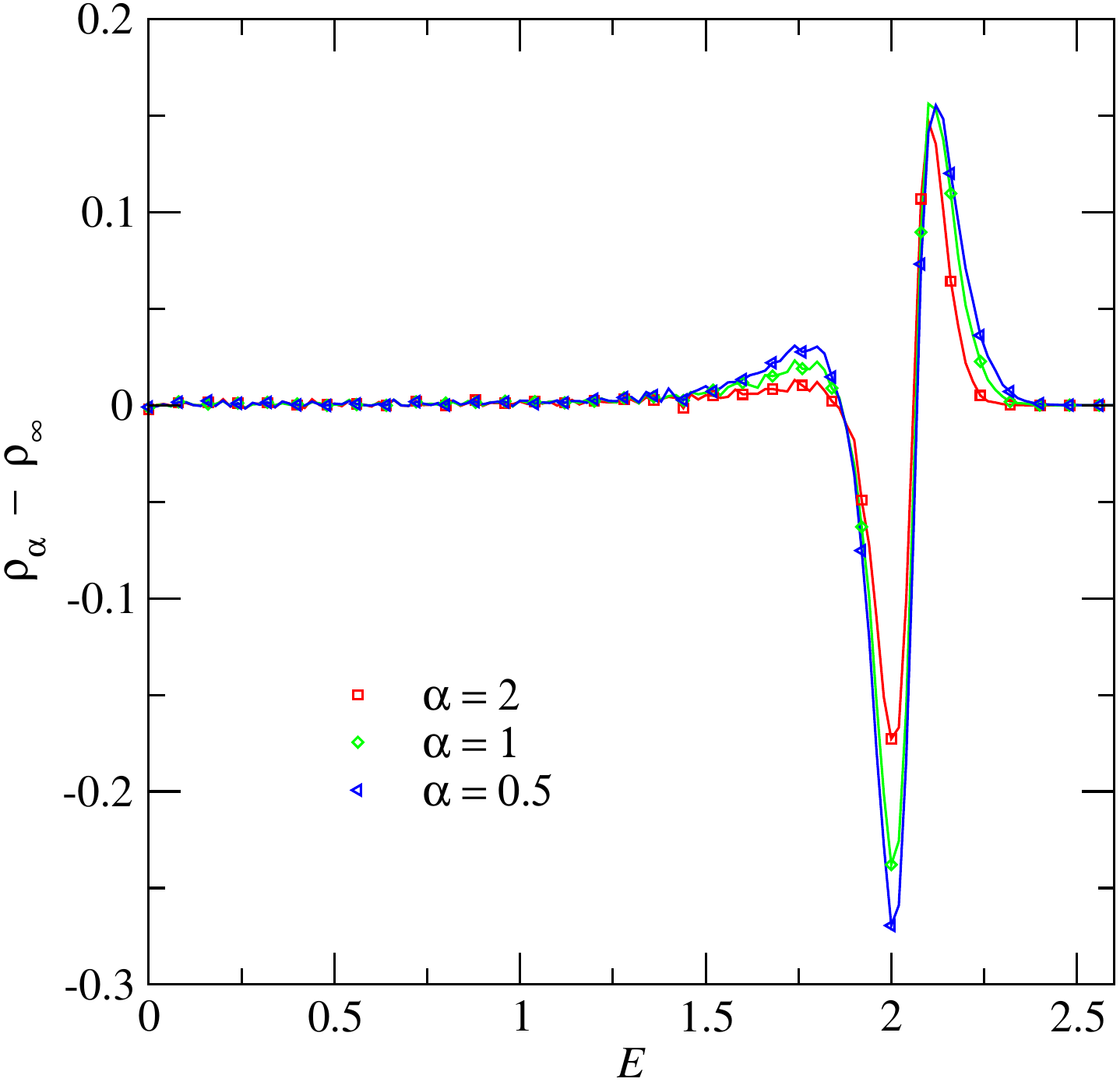}
	\caption{(Color online) Difference of the DOS $\rho_\alpha$
          for a 1D chain of length $L=2^{13}$ and $W=0.5$ compared to
          the uncorrelated case.  Different symbols/colors denote
          various correlation parameters $\alpha$ as given in the
          legend.}
	\label{fig:1DDOS}
\end{figure} 
Another important quantity for transport studies in general is the DOS
$\rho(E)$.  For an ordered system the DOS can be calculated
analytically.  In this case it develops van Hove singularities at the
band edges $|E|=2$, which is typical for 1D systems \cite{Eco06} and
indicates the presence of long-range order \cite{KraM93}. In case of
uncorrelated on-site energies the van Hove singularities are smeared
out while maintaining peaks at $|E|=2$.  Inside the unperturbed band
the DOS is only weakly changed compared to the DOS of the ordered
system. For $|E|>2$ there is an exponential band tail \cite{KraM93}.
For correlated on-site energies Fig.\ \ref{fig:1DDOS} illustrates that
the van Hove peak is even less pronounced and shifted towards the band
center, which results in a decreased DOS at $|E|=2$ with decreasing
$\alpha$.  On the other hand, in the band tails the DOS is getting
larger for decreasing $\alpha$.

\begin{figure}[tb]
	\center
	\includegraphics[width=\columnwidth]{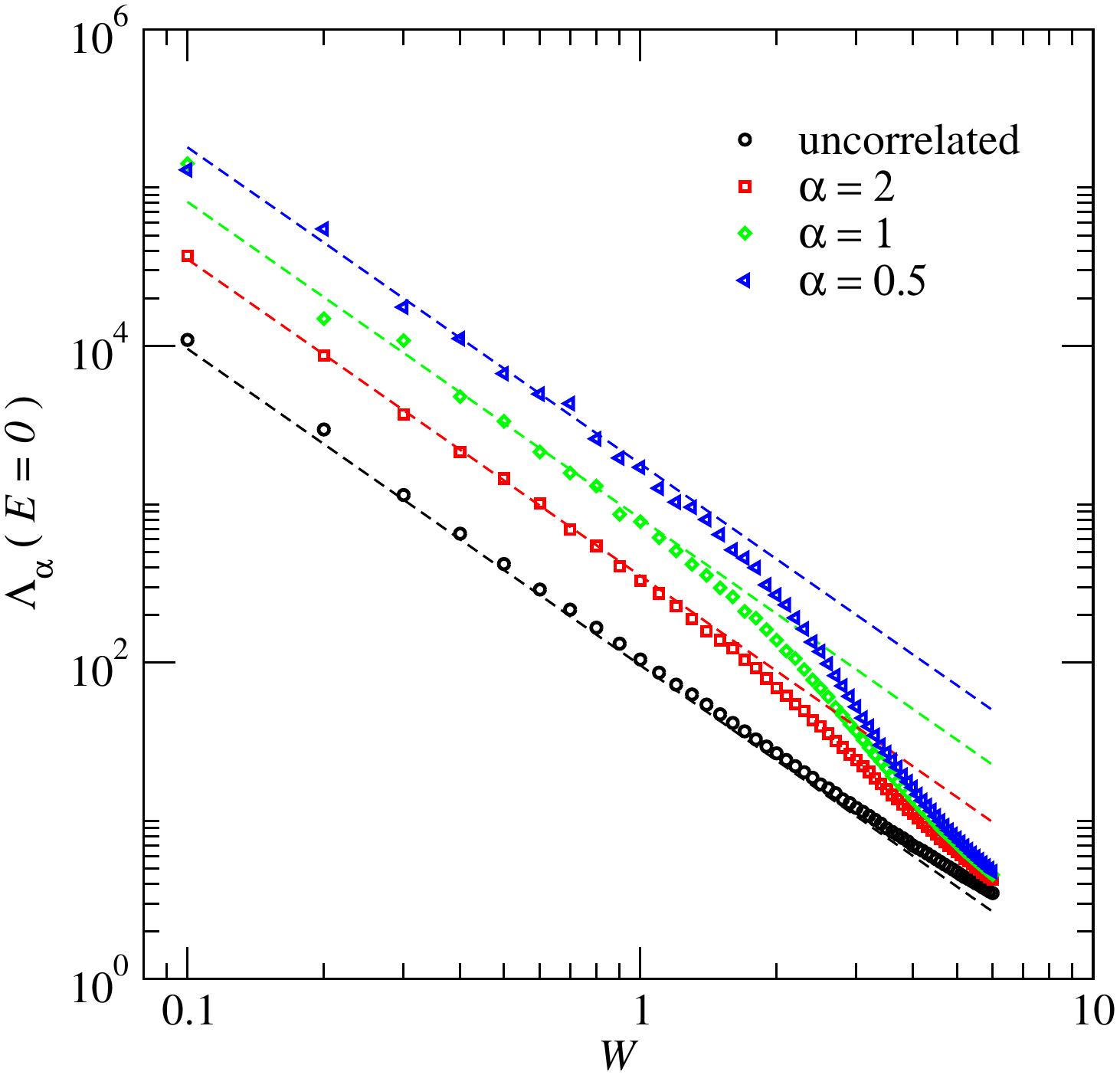}
	\caption{(Color online) Localization length $\Lambda_\alpha$
          vs disorder strength $W$ at the band center $E=0$ for a 1D
          chain of length $L=2^{19}$ obtained from TMM calculations.
          Different symbols denote various correlation parameters
          $\alpha$.  Dashed lines show $\Lambda_\alpha$ given by Eq.\
          \eqref{eq:1DWeakDisExpansion}.}
	\label{fig:1DDisLocE0}
\end{figure}

In the following we discuss the two regions separately and in more
detail.  Firstly, we focus specifically on the behavior in the center
of the band.  Figure \ref{fig:1DDisLocE0} shows the disorder strength
dependence of the localization length at $E=0$. One can see that
introducing correlations leads to a systematically larger localization
length. Only for very strong disorder the influence of the
correlations vanishes and the localization length for finite $\alpha$
approaches the uncorrelated localization length. Also shown is the
localization length obtained from Eq.\ \eqref{eq:1DWeakDisExpansion}
using the same parameters as in the TMM calculation. It is important
to notice, that there is no fitting procedure involved at this point.
Therefore, Eq.\ \eqref{eq:1DWeakDisExpansion} allows for an {\it
  independent} check of the numerics and in particular of the correct
behavior of the generated correlations.  Figure \ref{fig:1DDisLocE0}
shows a good agreement of $\Lambda$ and the weak disorder expansion
result \eqref{eq:1DWeakDisExpansion} for $W < 1$.
In contrast to $\Lambda$ the DOS in the band center has no particular
dependence on the correlation strength for weak disorder as
demonstrated in Fig. \ref{fig:1DDOS}. Therefore the influence of the
long-ranged correlations can be interpreted in accordance with Eq.\
\eqref{eq:1DWeakDisExpansion} as an effective disorder of the
potential. Hereby, the effective disorder strength becomes smaller
with decreasing $\alpha$.

\begin{figure}[bt]
	\center
	\includegraphics[width=\columnwidth]{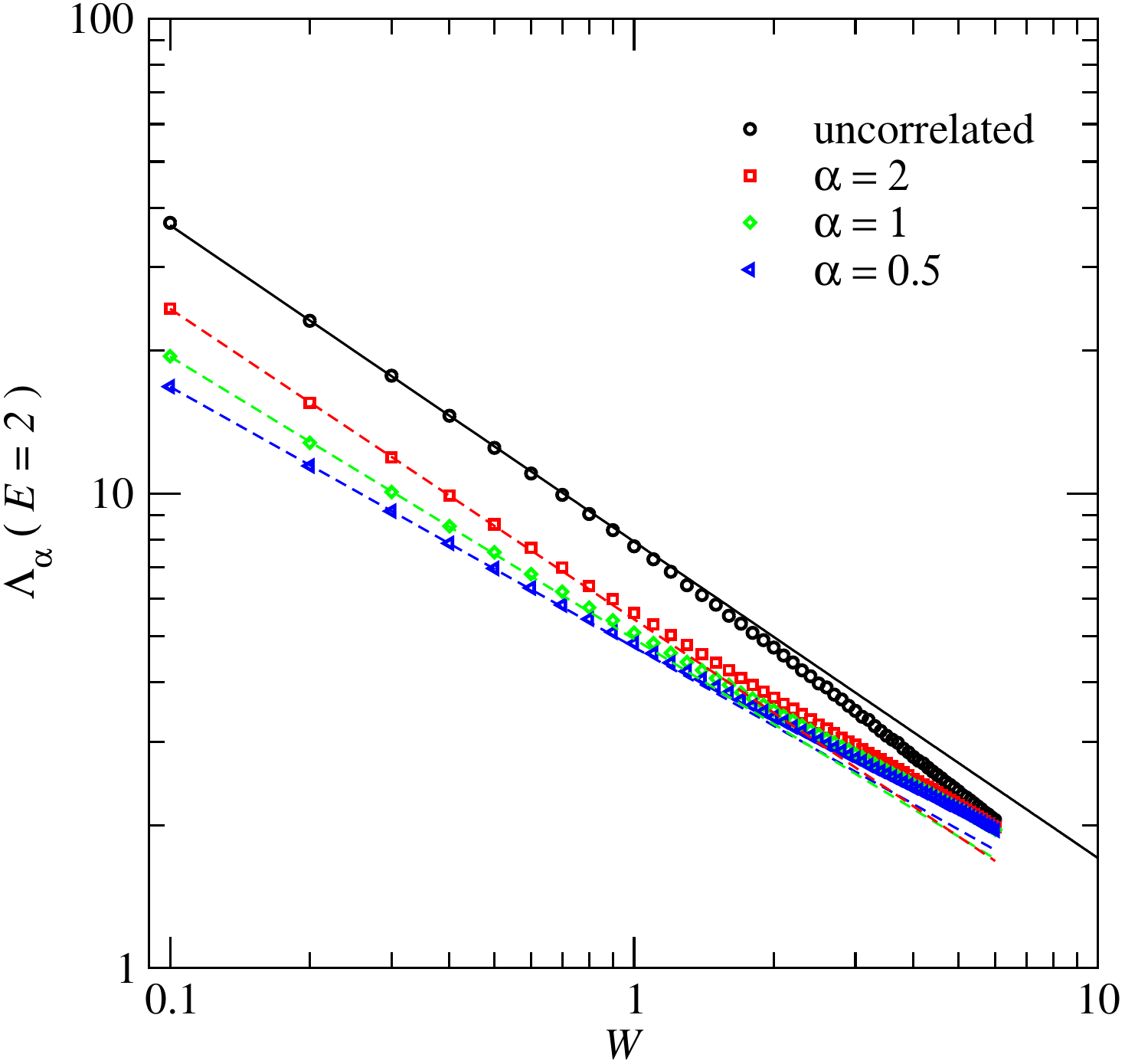}	
	\caption{(Color online) Localization length $\Lambda_\alpha$
          vs disorder strength $W$ at the unperturbed band edge $E=2$
          for a 1D chain of length $L=2^{19}$.  Different symbols
          indicate results obtained from TMM calculations.  Error bars
          are smaller than the symbol size.  The full line shows the
          exact result for uncorrelated disorder \cite{DerG84}.
          Different colors denote various correlation parameters
          $\alpha$. Dashed lines show results of a least squares fit
          of $\Lambda_\alpha \propto W^{-y}$ to the data.  }
	\label{fig:1DDisLocE2}
\end{figure}

\begin{figure}[tb]
	\center
	\includegraphics[width=\columnwidth]{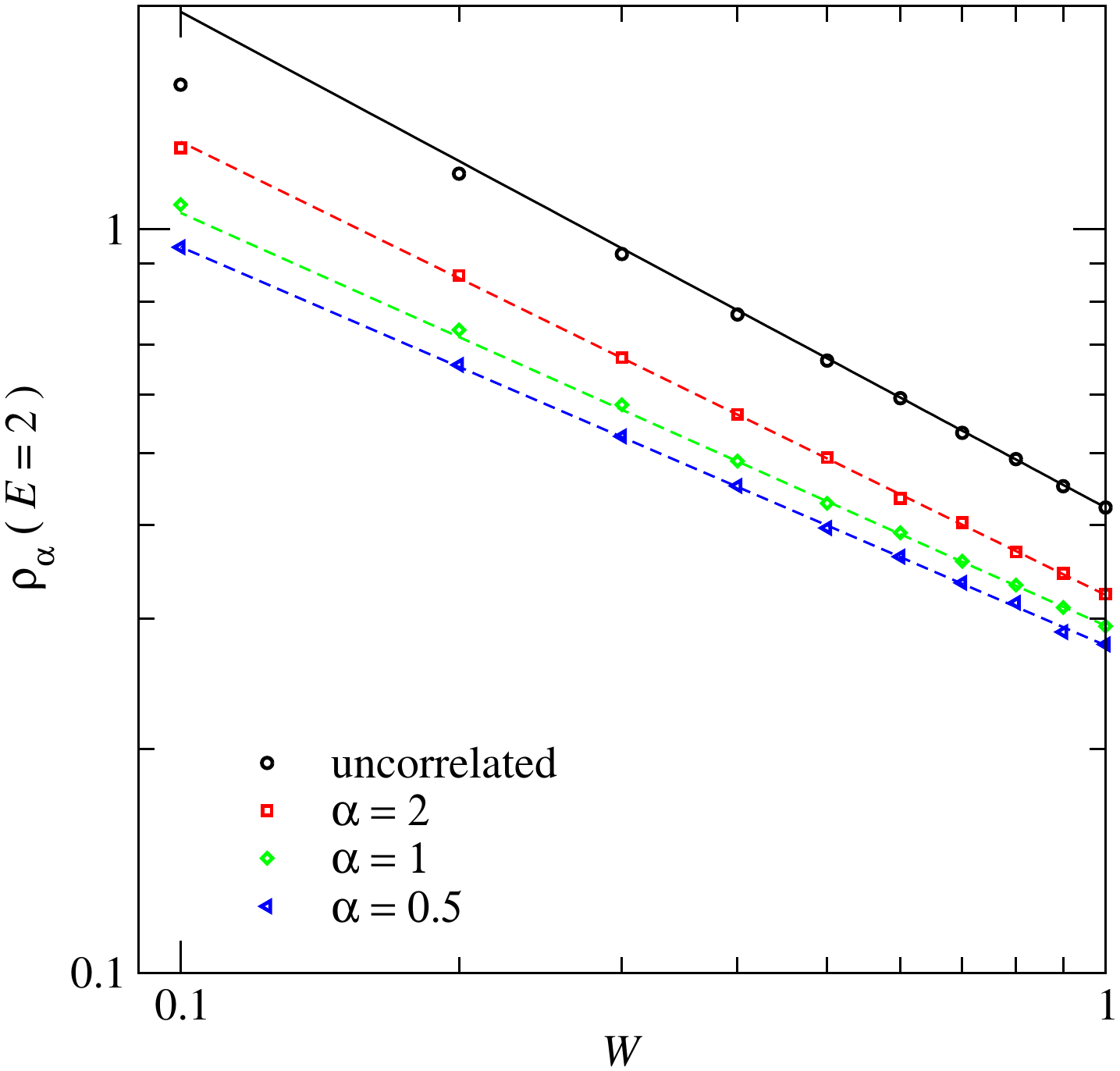}
	\caption{(Color online) DOS $\rho_\alpha$ vs disorder strength
          $W$ at the unperturbed band edge $E=2$ for a 1D chain of
          length $L=2^{13}$.  Different symbols indicate various
          correlation parameters $\alpha$.  The full line shows the
          exact result for uncorrelated disorder \cite{DerG84}.
          Dashed lines denote results of a least squares fit of
          $\rho_\alpha \propto W^{-y}$ to the data.}
	\label{fig:1DDisDOSE2}
\end{figure}

A qualitatively different behavior of the localization length is
observed for energies close to the unperturbed band edge ($|E|=2$).
This is shown in Fig.\ \ref{fig:1DDisLocE2}.  Here the localization is
enhanced in case of a correlated potential and $\Lambda$ is
consequently smaller compared to the uncorrelated case. However, for
weak disorder $W < 1$ one can still observe a power-law decay
according to Eq.\ \eqref{eq:1DPowerDecay}, but with an exponent
$y<2/3$. The results of least-squares fits of Eq.\
\eqref{eq:1DPowerDecay} to the numerical data are summarized in Tab.\
\ref{tab:1DDisLocFit} and also shown in Fig.\ \ref{fig:1DDisLocE2}.
The obtained exponents for the uncorrelated potential as well as for
$\alpha=2.0$ and $0.5$ agree with the values predicted by Eqs.\
\eqref{eq:1DPowerDecay}. Only for $\alpha=1.0$ we observe a distinct
deviation from the expected value $y=2/3$ which could be a consequence
of the modified FFM \cite{MakHSS96}. From Eq.\ \eqref{eq:SpecDens} one
expects the FFM to generate an uncorrelated potential for $\alpha=1$
since $S(q)$ becomes independent of $q$ in this case. However, due to
the short-range cutoff introduced in Eq.\ \eqref{eq:MakseCorrFun}, the
spectral density is given in terms of a modified Bessel function
\cite{MakHSS96}, which is independent of $q$ only in the limit
$\alpha\to\infty$.  Therefore, the behavior of the localization length
and the DOS at the band edge is sensitive to the specific choice of
the short-range cutoff for sufficiently weak disorder.
On the other hand, for very strong disorder the localization lengths
for the correlated cases coincide with the uncorrelated case.
Figure \ref{fig:1DDisDOSE2} shows the DOS at the band edge for small
disorder strengths.  Similar to the localization length one observes a
power-law decay of the DOS, which is also in accordance with the
scaling law \eqref{eq:BEScaling}. The fitted exponents are listed in
Tab.\ \ref{tab:1DDisLocFit}. Their values agree with the results
obtained from the localization length but are less accurate.
\begin{table}[pb!]
  \caption{Results of a least squares fit of $\Lambda_\alpha \propto W^{-y_\Lambda}$ and 
    $\rho_\alpha \propto W^{-y_\rho}$ to the numerical data for weak disorder. For
    $\Lambda_\alpha$ and $\rho_\alpha$ the range $W\in [0.1, 0.5]$ and $W\in [0.4,1.0]$ was used, respectively.
    The obtained exponents
    are compared to the expected values $y$ according to Eqs.\ \eqref{eq:1DPowerDecay}.}
	\begin{tabular}{lcccc}
	 \hline	
	 \hline
	 $\alpha$		& $\infty$ 			& $2$ 					& $1$ 					& $0.5$ \\
	 \hline
	 $y$				& $2/3\approx0.67$ & $2/3\approx0.67$ & $2/3\approx0.67$ & $4/7\approx0.57$\\
	 \hline
	 $y_\Lambda$	& $0.68\pm0.01$ 	& $0.66\pm0.01$ 	& $0.60\pm0.02$	& $0.55\pm0.01$ \\
	 $y_\rho$		& $0.65\pm0.01$		& $0.61\pm0.03$	& $0.56\pm0.02$	& $0.54\pm0.03$\\
	 \hline	
	 \hline
	\end{tabular}
  \label{tab:1DDisLocFit}
\end{table}

%%%%%%%%%%%%%%%%%%%%%%%%%%%%%%%%%%%%%%%%%%%%%%%%%%%%%%%%%%%%%%%%%%%%%%%%%%%%%%%%%%%%%%%%%%%%%%%%%%%%%%%%%%%%%%%%%%%%%%%%%%%%%%%%%
%
% CHAPTER: Conclusions
%
%%%%%%%%%%%%%%%%%%%%%%%%%%%%%%%%%%%%%%%%%%%%%%%%%%%%%%%%%%%%%%%%%%%%%%%%%%%%%%%%%%%%%%%%%%%%%%%%%%%%%%%%%%%%%%%%%%%%%%%%%%%%%%%%%
%
\section{Summary and Conclusions}
\label{sec:summary}
In summary, we have numerically investigated the role of long-range
correlated disorder on Anderson localization in 1D systems.
Specifically, we have studied the behavior of the localization length
and the DOS as functions of energy and disorder strength for different
correlation exponents. Hereby, we found two regions in the electronic
band, where universal behavior can be observed. In the vicinity of the
band center and for weak disorder the localization length is
determined by an effective disorder strength. The localization length
remains proportional to $W^{-2}$ in this regime. The explicit
expression, which has been derived in Ref.\ \cite{IzrK99}, is given by
Eq.\ \eqref{eq:1DWeakDisExpansion}.
A qualitatively different behavior is found at the band edge. Here,
the localization length and DOS are given by Eqs.\
\eqref{eq:BEScaling}, which implies a power-law behavior for both
quantities as functions of disorder strength.  A comparison with the
WNM suggests that in this region the long-range correlations lead to
an effectively reduced dimension.

Overall, we find good agreement of our numerical results with the
respective analytical expressions. However, we also find that the
particular realization of the long-range correlations plays an
important role. In the present case the FFM \cite{MakHSS96} makes use
of a modified correlation function given by Eq.\
\eqref{eq:MakseCorrFun}, which contains the desired long-range but
also short-range contributions. Our numerical results indicate that
the modification leads to observing an unexpected influence of the
correlations even for correlation exponents $\alpha>1$. This behavior
should be taken into account for investigations involving realistic
disorder potentials with long-range correlations.
%
%%%%%%%%%%%%%%%%%%%%%%%%%%%%%%%%%%%%%%%%%%%%%%%%%%%%%%%%%%%%%%%%%%%%%%%%%%%%%%%%%%%%%%%%%%%%%%%%%%%%%%%%%%%%%%%%%%%%%%%%%%%%%%%%%
%
% CHAPTER: Bibliography
%
%%%%%%%%%%%%%%%%%%%%%%%%%%%%%%%%%%%%%%%%%%%%%%%%%%%%%%%%%%%%%%%%%%%%%%%%%%%%%%%%%%%%%%%%%%%%%%%%%%%%%%%%%%%%%%%%%%%%%%%%%%%%%%%%%
%

\end{document}